# Concept Based vs. Pseudo Relevance Feedback Performance Evaluation for Information Retrieval System

Mohammed El Amine Abderrahim
University of Tlemcen, Faculty of Technology
Laboratory of Arabic Natural Language Processing
BP 230 Chetouane, Tlemcen, Algeria

**Abstract**

*This article evaluates the performance of two techniques for query reformulation in a system for information retrieval, namely, the concept based and the pseudo relevance feedback reformulation.*

*The experiments performed on a corpus of Arabic text have allowed us to compare the contribution of these two reformulation techniques in improving the performance of an information retrieval system for Arabic texts.*

**Keywords:** *Arabic NLP, Arabic Information Retrieval, Query Expansion, Arabic WordNet,* pseudo relevance feedback.

## 1. Introduction

In Information Retrieval System (IRS), researchers work distinguishes two approaches for the Query Reformulation (QR). The first approach is called direct reformulation and consists of adding new terms to the initial query, it is based on external resources such as ontologies and thesaurus or the relation of co-occurrences between terms, i.e., Concept Based (CB) query reformulation [Baeza-Yates and Berthier, 1999, Carpineto and Romano, 2012, Efthimiadis, 1996]. Whereas, the second approach, is called indirect reformulation and consists of modifying the user query by taking into account a list of documents already considered as selected and all the terms added to the query come only from the documents' collection and not from an external resource [Baeza-Yates and Berthier, 1999, Carpineto and Romano, 2012, Salton and Buckley, 1990]. This process is called Relevance Feedback (RF) once supervised and Pseudo Relevance Feedback (PRF) once automatic.

According to [Abderrahim and A, 2010, Harb et al., 2011, Xu et al., 2002, Baziz, 2005, Nathalie et al., 2008, Wedyan et al., 2012], the CB QR has a positive effect in information retrieval. In this same context, the studies applied in [Abderrahim and A, 2012, Xu and Croft, 2000, Kanaan et al., 2005, Salton and Buckley, 1990, Abderrahim et al., 2013] show that the RF (or PRF) is simpler to be realized and allows an improvement of IRS performances.

This paper aims to evaluate and compare the contribution of these two reformulation techniques (CB and PRF) in improving IRS performance for Arabic texts.

## 2. Concept Based (CB) and Pseudo Relevance Feedback (PRF) query reformulation

Reformulating a query is an attempt to improve user queries by: adding terms; re-weight existing terms; removing terms; or a combination of those methods. In this section, we examine two approaches for improving the initial query formulation through query expansion. These approaches are automatic (i.e., user intervention is not required at any stage other than the formulation of the original query) and grouped in two main categories: (a) approaches based on information derived from knowledge structures such as thesauri (also called CB direct reformulation); and (b) approaches based on information derived from the set of documents initially retrieved (also called PRF indirect reformulation). In these approaches, an original query is run using conventional information retrieval techniques. Then, related terms are extracted from the top "n" (generally n=10) documents (also called local document set) that are returned in response to the original query. The related terms are then added to the original query, and the expanded query is run again to return a new set of documents, which are returned to the user. In the literature, two techniques are proposed for the extraction of terms from de the local document set [Baeza-Yates and Berthier, 1999]:

(a) The local clustering: this consists of constructing a matrix of association that quantifies the correlation relations of terms got from the set of documents that were returned in response to the initial query. According to the method of construction of the correlation relations, we notice three types of clusters: association clusters, metric clusters and scalar clusters.

(b) The local context analysis: it consists of using the concepts instead of the keywords to represent the document [Xu and Croft, 2000].

As mentioned above, PRF strategies are based on expanding the query with terms correlated to the query terms. Such correlated terms are those present in local clusters build from the local document set. The idea of local clustering approach is to build global structures such as association matrices which quantify term correlation and to use correlated terms for query expansion.

When using knowledge structures, expansion terms are determined from pre-fabricated term dependency matrices, and no significant work has to be done during query time to expand queries. During query time, queries are expanded simply by looking up related terms in the appropriate knowledge structures.

The CB QR (see figure 1) consists of analyzing the query aiming at detecting the terms which correspond to concepts in the external resource (Arabic WordNet (AWN) in our case). These terms are replaced by the proximate concepts by using the AWN semantic relations (only the synonymy relation is tested in our work).

The basic procedure of the CB QR is:
- The user issues a query.
- The analyzer extracts terms which possibly correspond to concepts from the initial query.
- For each term extracted, the system try to selects all it proximate concepts from AWN (in our case, proximate concepts of a term "t" are synonyms of t from AWN).
- The query is then expanded with all proximate concepts.
- The system displays a set of retrieval results.

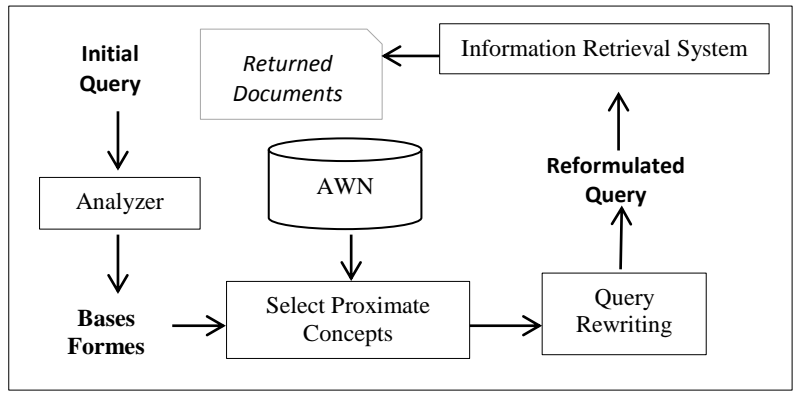

Figure 1 The CB query reformulation

The idea behind the CB QR is that for each term "t" in a query, the query can be automatically expanded with synonyms of "t" from AWN.

The PRF is a technique that consists of modifying the initial query of the user by adding some terms got from the list of documents retrieved in the IRS. The basic procedure of the PRF is:
- The user issues a query.
- The system returns an initial set of retrieval results.
- The system selects the "n" first documents as relevant.
- The system computes a better representation (T terms) of the information need based on the "D" documents (also called local document set).
- The query is then expanded with all extracted terms.
- The system displays a revised set of retrieval results.

The PRF is based on three steps (see figure 2):
- The samples: it consists of selecting a set of «D» documents (samples) among the returned ones by the IRS and judged as pertinent.
- The extraction of evidences: it consists of establishing the list of «T» terms judged pertinent for the expansion of the query.
- The rewriting of the query: It consists of enriching the query with terms found in the previous step.

According to the analysis of D and T for Arabic text in [Abderrahim et al., 2013], we have used D=15 and T=7.

The idea behind the PRF QR is that the documents retrieved for a given query "q" are examined at query time to determine terms for QE.

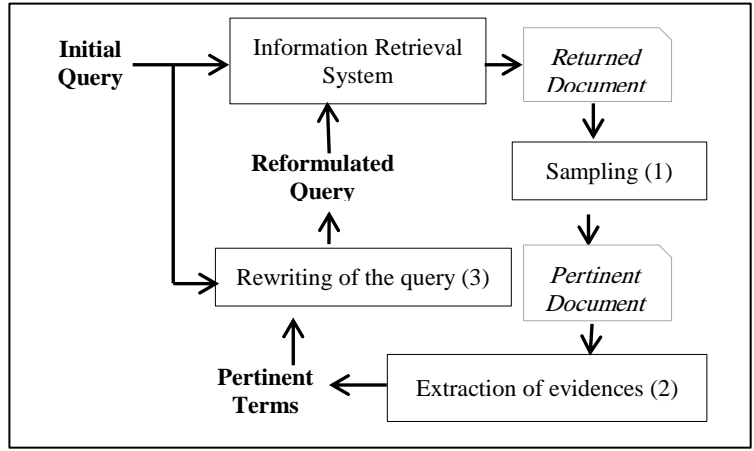

Figure 2 The PRF query reformulation

In the following description of experimentation and discussion of obtained results are displayed.

## 3. Experiments' description

An external resource for research similar concepts in the phase of QR is used in our experiment. This resource is AWN, it is one of the most important lexical database of the Arabic language following the development process of Princeton English WordNet (PWN) [Fellbaum, 1998] and EuroWordNet [Vossen, 2004].

Like EuroWordNet, AWN is developed in two phases by:
- Building a core wordnet around the most important concepts (Base Concepts). This core is highly compatible with wordnets in other languages that are developed according to the same approach.
- Extending the core wordnet to more specific concepts.

In AWN, nouns, verbs, adjectives, and adverbs are organized into synonymous words, called synsets. AWN consists of 11 269 synsets: 7961 nominal, 2538 verbal, 661 adjectival, and 110 adverbial [Alkhalifa, 2006, Elkateb et al., 2006, Black et al., 2006, Rodríguez et al., 2008(a)].

The AWN system falls into two parts: the AWN lexical database; and the software tools used to access the database. Two different web-based interfaces have been developed for AWN: A lexicographer's web interface and a user's web interface. The first one is designed to support the task of adding, modifying, moving or deleting AWN synsets. The second one enables the user to consult AWN and search for Arabic words, Arabic roots, Arabic synsets, English words, synsets offsets for English PWN.

The AWN Browser is a stand-alone Java application, it include browsing AWN, searching for concepts, and updating AWN with data from the lexicographers. Searching can be done using either English or Arabic. In Arabic, the search can be carried out using either Arabic script or Buckwalter transliteration and can be for a word or root form, with the possibly use of diacritics. The browser allows a user to navigate via hyponym and hypernym relations between synsets of AWN and PWN. It should be noted that the SUMO ontology navigation is also integrated into the browser. Users will be able to search or browse AWN using SUMO as the interlingual index between English and Arabic. The browser is freely available for downloading from Sourceforge at: http://www.globalwordnet.org /AWN/ AWNBrowser.html.

The database structure comprises several entity types: item, word, form, link, and authorship. Item are conceptual entities that contain information about synsets, both English and Arabic. An item has a unique identifier. A word entity holds information about words within synsets, both English and Arabic. A form is an entity that contains lexical information about different forms of Arabic words. The forms are root and/or the broken plural form, where applicable. A link relates two items; this can be within a particular language such as "hyponymy" or between equivalent Arabic and English synsets. The authorship entity holds information about when and by whom a synset, word or form was created.

The data model of AWN is implemented in a MySQL database and can be exported to other format such as XML or CSV.

Currently, AWN is still under construction and efforts are underway to partially automate the process using bilingual Arabic-English resources, PWN and applying morphological rules such in [Rodríguez et al., 2008(b), Alkhalifa et al., 2009, Cavalli-Sforza, 2013, Abouenour, 2013].

A corpus is also used in our experiment. It is composed of more than 22,000 documents containing Arabic texts from different fields. The main characteristics of this corpus can be found in Table 1.

| **Number of text files** | 22 428 |
|---|---|
| **Fields** | Health, sports, politics, sciences, religion, astronomy, nutrition, law, tales, family |
| **Size** | 180 MB |
| **Number of words** | 17 000 000 |
| **Number of terms** | 193 736 |

Table 1: The main features of the used corpus

To carry out our experimentation the freely available API Lucene search engine is used (http://lucene.apache.org), were the whole of the documents of our corpus is indexed and resulted in one indexes of approximately 37 MB. The process of CB and PRF query reformulation are developed in Java language.

Three strategies are planned for information retrieval:
- Simple Retrieval or without reformulation (SR): A set of 50 queries (see table 2) is used.
- Concept Based Retrieval (CBR): A set of 50 queries deduced from the initial queries by a blind enrichment is used. The set of the synonyms found in AWN is added to the initial query (see table 2). The following algorithm is implemented for this strategy.

```
// Algorithm of CB query reformulation
Begin
    For each query q_i  (i=1,50)
        1. Extraction of evidences
            For each term t_j of the query q_i
                Find the list S_j of t_j synonyms from AWN
                    S_j = AWN_Synonyms (t_j)
        2. Rewriting the query
                Construct the new query:  q_new = q_i U S_j
End
```

Table 2 Examples of queries for CBR and PRFR

| Query N° | Simple query (SR) | New enriched query (CBR) | New enriched query (PRFR) |
|---|---|---|---|
| 1 | سعر النفط (Price of oil) | سعر النفط (Price of oil) ، قيمة مالية (Financial Value) ، تكلفة (Cost) ، ثمن (Price) ، سعر الوحدة (Unit Price)، تعريفة (Tariff) ، وحدة حرارية (Thermal Unit) ، سعرة (Calories) | سعر النفط (Price of oil) ، امريكي (American)، برميل (Barrel)، دول (States)، اسعار (Prices)، انتاج (Production)، اوبك (OPEC) |
| 2 | نظام رأسمالي (Capitalist System) | نظام رأسمالي (Capitalist System) ، انتظام (Uniformity)، حالة نظامية (Regular situation)، منظومة (Système)، برنامج (Program) ، خطة (Plan) | نظام رأسمالي (Capitalist System) ، بنوك (Banks)، عالم (World)، ازمة (Crisis)، حكومة (Government) |

- PRF Retrieval (PRFR): A set of 50 queries deduced from the initial queries by PRF enrichment is used (see table 2). The following algorithm is implemented for this strategy. It implements the local clustering technique to extract the most pertinent «T» terms that serve in the reformulation of the initial query.

```
// Algorithm of the PRF QR
Begin
    For each query qi (i=1, 50) do

    1. Interrogation of the collection of documents
    2. Sampling : select the «D=15» first returned documents: D_F
    3. Extraction of evidences
       Construct the matrix of local association (term - term) from the set of distinct terms of
       D_F : S⃗  with each element
                    S_{u,v} = Σ_{d_j ∈ D_F} f_{S_u,j} × f_{S_v,j}

                f_{S_u,j} : represents the term frequency   S_u in the document  d_j

                S_{u,v} : expresses the correlation between u and v
       For each term t of  q_i extract its local association clustering C_i set from the « T=7 »
       highest values  S_{u,v}  (v ≠ u) of the u^th line of S⃗
    4. Rewriting of the query : construct the new query : q_new = q_i U C_i
End
```

The retrieval results obtained by these various types of queries are saved in various files and various values of recalls and precisions of the system are calculated for each type of retrieval and query.

## 4. Analysis and discussion of the results

### 4.1 The precision at different level of documents

Figure 3 presents the precision obtained at 5, 10, 20 and 100 documents (P@5, P@10, P@20, P@100) from different types of reformulation, and shows that the use of AWN did not improve the precision at different levels of documents after the QR. Moreover, the reformulation made by PRFR is much better than CBR.

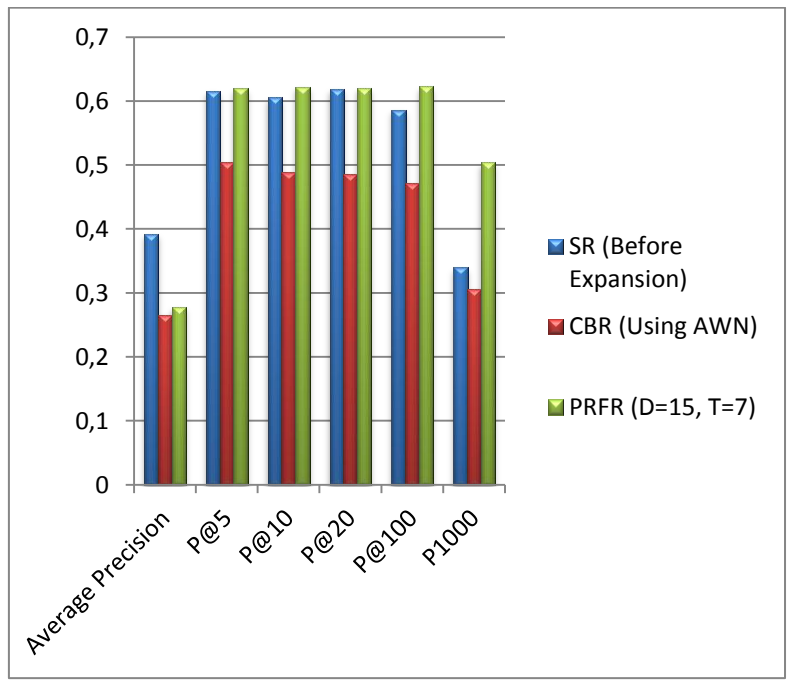

Figure 3 Comparison of the reformulation results

### 4.2 The precision at different level of recall

Table 3 presents the precisions at 11 levels of recall for each retrieval type.

| Recall | SR | CBR | PRFR |
|---|---|---|---|
| **0** | 0,620 | 0,484 | 0,628 |
| **0,1** | 0,616 | 0,453 | 0,625 |
| **0,2** | 0,587 | 0,427 | 0,594 |
| **0,3** | 0,557 | 0,392 | 0,566 |
| **0,4** | 0,543 | 0,359 | 0,546 |
| **0,5** | 0,508 | 0,330 | 0,527 |
| **0,6** | 0,464 | 0,305 | 0,506 |
| **0,7** | 0,431 | 0,293 | 0,475 |
| **0,8** | 0,402 | 0,281 | 0,426 |
| **0,9** | 0,368 | 0,257 | 0,321 |
| **1** | 0,304 | 0,207 | 0,190 |

Table 3 The precisions at 11 levels of recall for each retrieval type

Figure 4 shows the curves' recall / precision obtained from table 4.

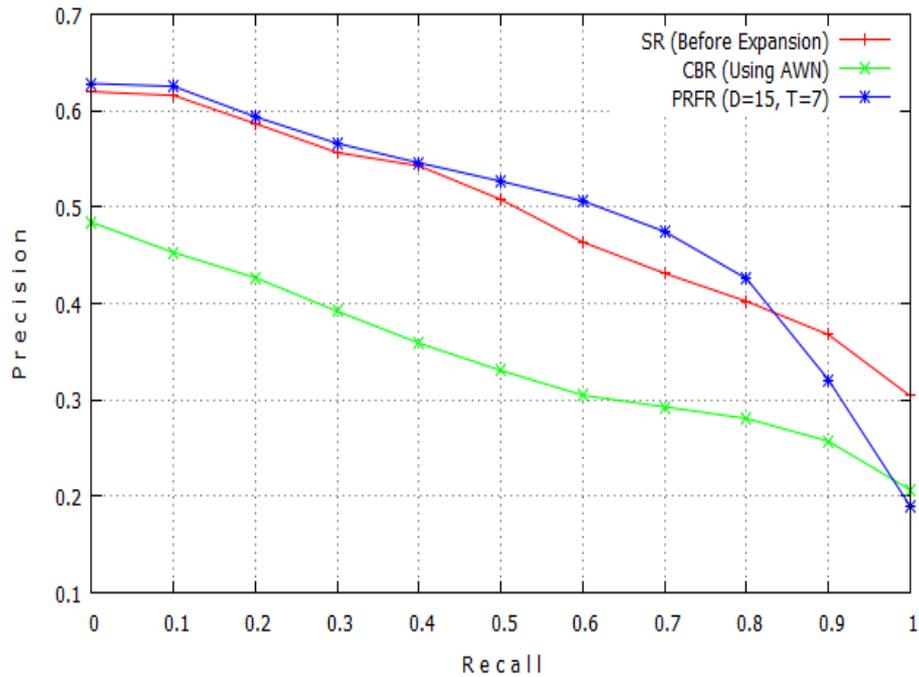

Figure 4 Comparison between recall / precision curves for each retrieval type

From figure 4 we can deduce that:
- The PRFR has led to an overall improvement in the recall interval [0, 0.8].
- CBR has not led to an overall improvement in the performance of the Arabic IRS.
- The PRFR is more beneficial than CBR.

To understand the effect of different types of retrieval on each query, various measures that are mainly based on the comparison of results before and after expansion have been established.

For a given query, three cases can arise and are as follows:
- Improvement (+): All precisions (at 11 points of recall) before are lower than those after. In other words, the curve (recall / precision) after is over before.
- No improvement (-): is the inverse of the previous case. The curve (recall / precision) before is over after.
- No decision (X): for some precisions, there is an improvement but for others there is no improvement. In other words, there is an intersection of the two curves (recall / precision).

Table 4 shows for each query used in the experiment the indicator: Improvement (+), No improvement (-) or No decision (X).

| Query N° | CBR | PRFR | Query N° | CBR | PRFR |
|---|---|---|---|---|---|
| 1 | - | - | 26 | - | + |
| 2 | X | + | 27 | + | + |
| 3 | - | - | 28 | + | - |
| 4 | X | + | 29 | - | X |
| 5 | X | - | 30 | - | X |
| 6 | - | X | 31 | - | X |
| 7 | - | X | 32 | + | - |
| 8 | - | X | 33 | - | X |
| 9 | + | - | 34 | - | X |
| 10 | - | X | 35 | - | + |
| 11 | X | X | 36 | X | + |
| 12 | + | + | 37 | - | - |
| 13 | X | - | 38 | + | X |
| 14 | X | X | 39 | - | - |
| 15 | - | X | 40 | - | X |
| 16 | - | X | 41 | - | X |
| 17 | - | - | 42 | - | X |
| 18 | - | X | 43 | X | X |
| 19 | X | X | 44 | X | + |
| 20 | - | X | 45 | - | X |
| 21 | X | X | 46 | + | - |
| 22 | X | X | 47 | - | X |
| 23 | X | + | 48 | - | X |
| 24 | X | X | 49 | - | - |
| 25 | - | - | 50 | - | - |

Table 4 The indicator: Improvement (+), No improvement (-) or No decision (X), of each retrieval type.

From table 4 we can deduce that, whatever the retrieval type, we note that there is:
- Improvement (+) in only two queries (queries number: 12 and 27) (4%).
- No improvement (-) in 7 queries (14%).
- No decision (X) in 6 queries (12%).

- A set of 35 queries (70%) for which there is at least an improvement in 12 queries (24%). For the remaining 23 queries (46%), there is no improvement or indecision.

From the point of improvement view, the previous facts suggest that:
- The reformulation by the use of an external resource or the PRFR can improve the performance of an Arabic IRS about 4%.
- There are queries that are hardly improvable with the QR.

In order to determine the best retrieval type (CBR or PRFR) the number of queries for each retrieval type has been counted (see table 5).

|  | Number of queries | |
|---|---|---|
|  | **CBR** | **PRFR** |
| Improvement (+) | 7 (14%) | 9 (18%) |
| No improvement (-) | 29 (58%) | 14 (28%) |
| No decision (X) | 14 (28%) | 27 (54%) |

Table 5 The number of queries satisfying the conditions: Improvement (+), No improvement (-) or No decision (X), depending on the retrieval type.

Results in table 5 can show that the reformulation by PRFR has the best rate of improvement (18%). Furthermore, the analysis of the recall / precision curves (see figure 4) confirms this result.

Finally, we can conclude that the contribution of the use of an external resource in an Arabic IRS is about 14%. Moreover, it appears that PRFR is better than CBR.

## 5. Conclusion

In this article, CB and PRF, two techniques for QR in Arabic IRS are examined and compared. The results obtained, firstly, have confirmed that there are queries that are hardly improvable with the query reformulation and allowed measure the contribution (4%) of these two techniques in improving the overall performance of the Arabic IRS.

As for the queries' number that has really led to an improvement, comparison of results has showed that the PRF QR is better than CB QR. Moreover, in terms of recall / precision, the PRF QR is more beneficial than CB QR.

As for perspectives it can be said that this study has paved the way to test and compare other methods of reformulation with the same data of this experiment in order to determine the most appropriate technique to be adopted for Arabic IRS.

## 6. References

[Abderrahim and A, 2010] Abderrahim, M. E. A. and A, M.A. (2010). Using Arabic Wordnet for Query Expansion in Information Retrieval System. In IEEE The Third International Conference on Web and Information Technologies 16-19 June Marrakech Morocco.

[Abderrahim and A, 2012] Abderrahim, M.E.A. and A, M.A. (2012). Réinjection automatique de la pertinence pour la recherche d'informations dans les textes arabes. In IEEE 4th International Conference on Arabic Language Processing May 23 Rabat Morocco, CITALA, pages 77–81.

[Abderrahim et al., 2013] Abderrahim, M.E.A., Benameur, S., and Abderrahim, M.A. (2013). The number of terms and documents for pseudo- relevant feedback for ad- hoc information retrieval. International Journal of Computer Science Issues, IJCSI, 10(1):661–667.

[Abouenour, 2013] L. Abouenour· K. Bouzoubaa· P. Rosso, (2013). On the evaluation and improvement of Arabic WordNet coverage and usability. Lang Resources & Evaluation (2013) 47:891-917, DOI 10.1007/s10579-013-9237-0.

[Alkhalifa, 2006] Al khalifa, M. (2006). Arabic Wordnet and Arabic NLP. In Journées d'Etudes sur le Traitement Automatique de la Langue Arabe JETALA 5-7 June Rabat.

[Alkhalifa et al., 2009] Al Khalifa, M., & Rodrı́guez, H. (2009). Automatically extending NE coverage of Arabic WordNet using Wikipedia. InProceedings of the 3rd international conference on Arabic language processing CITALA'09, May, Rabat, Morocco.

[Baeza-Yates and Berthier, 1999] Baeza-Yates, R. and Berthier, R.-N. (1999). Modern Information Retrieval. Addison Wesley New York City NY ACM Press.

[Baziz, 2005] Baziz, M. (2005). Indexation conceptuelle guidée par ontologie pour la recherche d'information. PhD thesis, Université de Toulouse, Université Toulouse IIIPaul Sabatier.

[Black et al., 2006] Black, W., Elkateb, S., Rodriguez, H., Alkhalifa, M., Vossen, P., Pease, A., and Fellbaum, C. (2006). Introducing the Arabic WordNet Project. In Proceedings of the Third International WordNet Conference, pages 295–300.

[Cavalli-Sforza, 2013] Violetta Cavalli-Sforza, H. Saddiki, K. Bouzoubaa, L. Abouenour, M. Maamouri, E. Goshey Bootstrapping a WordNet for an Arabic Dialect from Other WordNets and Dictionary Resources, 10[th] ACS/IEEE International Conference on Computer Systems and Applications (AICCSA 2013), May 27-30, 2013, Fes/Ifrane, Morocco.

[Carpineto and Romano, 2012] Carpineto, C. and Romano, G. (2012). A survey of Automatic Query Expansion in Information Retrieval. ACM Computing Surveys (CSUR), 44(1):1.

[Efthimiadis, 1996] Efthimiadis, E. N. (1996). Query Expansion. Annual review of information science and technology, 31:121–187.

[Elkateb et al., 2006] Elkateb, S., Black, W., Vossen, P., Farwell, D., Rodríguez, H., Pease, A., and Alkhalifa, M. (2006). Arabic WordNet and the Challenges of Arabic. In Proceedings of Arabic NLP/MT Conference, London, UK. Citeseer.

[Fellbaum, 1998] Fellbaum, C. (ed.) (1998) WordNet: An Electronic Lexical Database. Cambridge, MA: MIT Press.

[Harb et al., 2011] Harb, H. M., Fouad, K. M., and Nagdy, N. M. (2011). Semantic Retrieval Approach for Web Documents. IJACSA) International Journal of Advanced Computer Science and Applications, 2(9):11–75.

[Kanaan et al., 2005] Kanaan, G., AlShalabi, R., AbuAlrub, M., and Rawashdeh, M. (2005). Relevance Feedback: Experimenting with a Simple Arabic Information Retrieval System with Evaluation. International Journal of Applied Science and Computations Vol 12 No 2 USA.

[Nathalie et al., 2008] Nathalie, H., Gilles, H., Josiane, M., and Bachelin, R. (2008). Ri et ontologies etat de lart. Technical report, INP Toulouse, Université Paul Sabatier Toulouse III.

[Rodríguez et al., 2008(a)] Rodríguez, R., Farwell, D., Farreres, J., Bertran, M., Alkhalifa, M., Martí, M.A., Black, W., Elkateb, S., Kirk, J., Pease, A., Vossen, P., and Fellbaum, C., (2008). Arabic WordNet: Current State and Future Extensions. Proceedings of The Fourth Global WordNet Conference, Szeged, Hungary. January 22-25, 2008.

[Rodríguez et al., 2008(b)] Rodríguez, H. Farwell, D. Farreres, J. Bertran, M. Alkhalifa, M. Martí M.A (2008) Arabic WordNet: Semi-automatic Extensions using Bayesian Inference. In Proceedings of the the 6th Conference on Language Resources and Evaluation LREC2008. Marrakech (Morocco), May 2008.

[Salton and Buckley, 1990] Salton, G. and Buckley, C. (1990). Improving Retrieval Performance by Relevance Feedback. Journal of the American Society for Information Science, 41(4):288–97.

[Vossen, 2004] Vossen P. (2004) EuroWordNet: a multilingual database of autonomous and language specific wordnets connected via an Inter-Lingual-Index. International Journal of Lexicography, Vol.17 No. 2, OUP, 161-173.

[Wedyan et al., 2012] Wedyan, M., Alhadidi, B., and Alrabea, A. (2012). The effect of using a thesaurus in arabic information retrieval system. International Journal of Computer Science Issues, IJCSI, 9(1):431–435.

[Xu and Croft, 2000] Xu, J. and Croft, W. B. (2000). Improving the Effectiveness of Information Retrieval with Local Context Analysis. ACM Transactions on Information Systems (TOIS), 18(1):79–112.

[Xu et al., 2002] Xu, J., Fraser, A., and Weischedel, R. (2002). Empirical Studies in Strategies for Arabic Retrieval. In Annual ACM Conference on Research and Development in Information Retrieval: Proceedings of the 25 th annual international ACM SIGIR conference on Research and development in information retrieval, volume 11, pages 269–274.